\documentclass[aps,prl,twocolumn,amsmath,showpacs,superscriptaddress]{revtex4-1}  
\usepackage{graphicx}  
\usepackage{bm}        
\usepackage{amssymb}   
\usepackage[caption=false]{subfig} 
\usepackage{units} 
\usepackage{epstopdf}

\begin{document}
\title{Exploiting lens aberrations to create electron vortex beams}

\author{L. Clark}
\email[Electronic address: ]{laura.clark@ua.ac.be}
\affiliation{EMAT, University of Antwerp, Groenenborgerlaan 171, 2020 Antwerp, Belgium}
\author{A. B\'{e}ch\'{e}} 
\affiliation{EMAT, University of Antwerp, Groenenborgerlaan 171, 2020 Antwerp, Belgium}
\author{G. Guzzinati}
\affiliation{EMAT, University of Antwerp, Groenenborgerlaan 171, 2020 Antwerp, Belgium}
\author{A. Lubk}
\affiliation{Triebenberglabor, University of Dresden, Zum Triebenberg 1, 01062 Dresden, Germany}
\author{M. Mazilu} 
\affiliation{SUPA, School of Physics and Astronomy, University of St Andrews, St Andrews, UK}
\author{R. Van Boxem}
\affiliation{EMAT, University of Antwerp, Groenenborgerlaan 171, 2020 Antwerp, Belgium}
\author{J. Verbeeck}
\affiliation{EMAT, University of Antwerp, Groenenborgerlaan 171, 2020 Antwerp, Belgium}

\date{\today}

\begin{abstract}
A model for a new electron vortex beam production method is proposed and experimentally demonstrated. 
The technique calls on the controlled manipulation of the degrees of freedom of the lens aberrations to achieve a helical phase front. 
These degrees of freedom are accessible by using the corrector lenses of a transmission electron microscope. 
The vortex beam is produced through a particular alignment of these lenses into a specifically designed astigmatic state and applying an annular aperture in the condensor plane.
Experimental results are found to be in good agreement with simulations.
\end{abstract}


\maketitle

Electron vortex beams are electron states with helical wave fronts, typically of the form $\Psi(r,\phi,z)=A(r)\exp(\mathrm{i} \ell \phi)\exp(\mathrm{i} k z)$, where $\ell \neq 0$  is called the topological charge of the beam~\cite{Bliokh2007}.
As such vortex beams propagate, the wave fronts spiral around the beam axis leading to a central phase singularity~\cite{OptAngMomBook}.
The resultant destructive interference leads to a topologically protected zero--intensity on--axis, creating a characteristic doughnut shape.
The probability current follows a helical path around the axis, and as such, there is an orbital angular momentum (OAM) around the beam axis~\cite{OptAngMomBook}.
In this case, the OAM  is quantized and proportional to $\ell$, with no theoretical upper limit, for non--apertured beams~\cite{Allen1992, zambrini}. 

Vortex beams are not a new phenomenon. The interplay of helical wave fronts and phase singularities was first considered in radio waves and ultrasound waves by Nye and Berry almost 40 years ago~\cite{Nye1974}.
This work broadened into the field of singular optics, with beams created to deliberately possess a vortex phase singularity, $\exp(\mathrm{i} \ell \phi)$, with the work of Vaughan and Willetts~\cite{VaughanWilletts} and Bazhenov \emph{et al. }~\cite{bazhenovscrew}.
The association of OAM and vortex beams was then discussed in a seminal paper by Allen \emph{et al.}~\cite{Allen1992}.

The field of optical vortex beam research is now very well established~\cite{BabikerNewBook, YaoOAM, OptAngMomBook}.
Furthermore, vortex beams have been demonstrated across a broad range of the electromagnetic spectrum~\cite{xrayvortex, radiovortex, irvortex, uvvortex} and the application of these has been considered in fields as diverse as astrophysics, quantum cryptography, biology and communications~\cite{astrovortex, twistedphotons, nanosurgery, vortexcommunication}.

However, vortex beams of matter waves were not considered in detail until 2007~\cite{Bliokh2007}, when electron vortex beams were first theoretically described (although, electron beams containing phase vortices had been discussed previously \cite{Allen2001}).
The first experimental demonstrations of electron vortex beams were presented in 2010~\cite{Uchida2010, VerbeeckNature}.
Uchida and Tonomura produced their vortex beams using an approximation to a stepped, spiral phase plate,
while Verbeeck \emph{et al.} used a holographic reconstruction method ~\cite{Uchida2010, VerbeeckNature}.

Subsequent research has since developed electron vortex beams much more broadly.
Electron vortex beams have been demonstrated for both detecting magnetic states within a material~\cite{VerbeeckNature, SchattBook} and as  a tool for manipulating nanoparticles~\cite{VerbeeckNanoManip}.
They have been produced with very high orders of OAM~\cite{McMorran2011, Saitoh2012}, and down to atomic size~\cite{VerbeeckAtomicScale, SchattSubNano} .

Alternative methods of production have also been developed.
The holographic reconstruction setup used by Verbeeck \emph{et al.}~\cite{VerbeeckNature} was adapted for application within scanning transmission electron microscopy ~\cite{VerbeeckSTEM}, while the previous work was better suited to conventional transmission electron microscopy.
Electron vortex production by mode conversion has also been demonstrated by Schattschneider \emph{et al.}~\cite{SchattPiPhasePlate}, based on the work by Allen \emph{et al.}~\cite{Allen1992}.

Very recently, electron vortex diffraction catastrophes were demonstrated by Petersen \emph{et al.} in the caustics resulting from highly aberrated electron probes~\cite{PetersenPRL}.
These results follow the optical theory as discussed by Berry and further demonstrate that the interplay between vortices and OAM is not always trivial~\cite{opticalcurrents}.
It is now established that there are many methods to produce both electron vortex beams, and  more broadly, electron beams carrying OAM.
Many of the experimental demonstrations of electron vortex phenomena are produced through methods adapted from the field of optical vortices~\cite{OptAngMomBook}.

We propose here, a fundamentally new method of producing electron vortex beams, through manipulation of the electron phase front using a multipole aberration corrector in a transmission electron microscope (TEM).
This can be considered as explicitly designing the shape of the electron wave such as to maximise the proportion of the beam in an OAM mode around the central axis.
We aim to adjust the values of the aberrations incident on the image plane within the TEM, such that when combined with an appropriate aperture, the electron beam in the microscope approximates the ideal $\Psi(r,\phi,z)=A(r)\exp(\mathrm{i} \ell \phi)\exp(\mathrm{i} k z)$.

In contrast with optical microscopy, the resolution of electron microscopes is limited by aberrations~\cite{WilliamsCarter}.
While in an idealised TEM, the diffraction limit is of the order of picometres, current resolution limits are around fifty times this value~\cite{Erni}.
This is due to both incoherent and coherent (geometric) aberrations. The incoherent aberrations occur due to factors such as mechanical vibrations and current stability. The geometric aberrations are due to deviations from ideal optics.

Multipole aberration correctors exist in many modern electron microscopes, to reduce the geometric aberrations inherently present due to cylindrically symmetric magnetic lenses~\cite{scherzer}.
Sets of multipole lenses, in sequence,  are able to apply an appropriate electromagnetic field configuration to counteract a particular deviation from an ideal system~\cite{rose2009}, by adjusting the phase of the incident electron wave.
Combinations of multipoles allow additional orders of aberration to be minimised~\cite{krivanek2008advances}, while inducing weaker, higher order aberrations~\cite{kirklandAber}.

The remaining geometric aberrations affect the wave--front by causing a position--dependent phase shift, wherein the symmetry of the aberration is related to its order.
Using our probe aberration corrector, it is currently possible to both measure, and adjust the values of the aberrations up to the $5^{th}$ order.
The ability to manipulate the phase of the electron wave in this way, and the current interest in electron vortex beams naturally leads to the investigation motivating this letter; to produce an electron vortex beam through use of aberration manipulation.

Following the notation of Saxton to define the aberration terms, the aberration function, up to fifth order,  can be expressed as~\cite{Saxton}:
\begin{align}
\chi=\frac{2\pi}{\lambda} &( A_0 \theta \cos ( \phi-\phi_{11} ) + \nonumber \\
\frac{1}{2} \theta^2 &( A_1 \cos ( 2 ( \phi - \phi_{22} ) ) +C_1 ) +\nonumber \\
\frac{1}{3} \theta^3 &( A_2 \cos ( 3 ( \phi - \phi_{33}) ) +B_2 \cos ( \phi - \phi_{31} )  ) + \nonumber \\
\frac{1}{4} \theta^4 &( A_3 \cos ( 4 ( \phi - \phi_{44}) ) +S_3 \cos ( 2( \phi - \phi_{42} )  ) + C_3 ) +\nonumber \\
\frac{1}{5} \theta^5 &( A_4 \cos ( 5 ( \phi - \phi_{55}) ) +B_4 \cos ( \phi - \phi_{51} )  +  \nonumber \\
&\qquad \qquad \qquad \qquad \quad  \; \; D_4 \cos ( 3 ( \phi - \phi_{53} ) ) )) \label{aberchi}
\end{align}
where $\chi$ is the phase shift on the electron wave front due to the aberrations at radial position, $\theta$,
 and azimuthal position, $\phi$.
The aberration parameters are: image shift, $A_0$; the orders of astigmatism, $A_{i\geq 1}$; coma, $B_i$; defocus, $C_1$; spherical aberration, $C_3$; star aberration, $S_i$; and three--lobe aberration $D_4$.
The $\phi_{ij}$ describe the relative angles of each aberration.
Higher order aberrations may be present, but they are not considered here as the effects are increasingly small, for small values of $\theta$.

To be able to manipulate the aberrations, $\exp ( - \mathrm{i} \chi )$, towards an approximation of an ideal phase vortex, $\exp ( \mathrm{i} \ell \phi )$, we consider the vortex phase variation, $\phi$, against electron phase, $\chi$.
The ideal phase function for an $\ell=1$ beam, is a linear variation of $\chi$, with $\phi$ increasing from $- \pi$ to $\pi$ as the azimuthal angle travels around a $2 \pi$ circuit. 
This can be visualised as a sawtooth phase with period of $2 \pi$, such that for an $\ell=1$ vortex, $\chi=-\phi$.

The determination of aberration parameters, leading to a vortex creation, requires the expansion of the sawtooth phase in a Fourier series:
\begin{equation}
\chi_{n}=\sum\limits_{n=1}^\infty  \frac{2}{n}(-1)^{n+1}\sin(n \phi) \label{sawtoothfouriereqn}
\end{equation}

 \begin{figure}
 \includegraphics[width=0.8\linewidth]{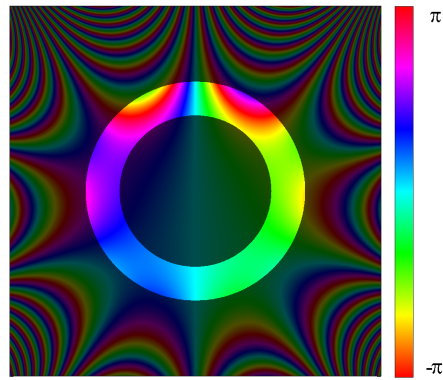}%
 \caption{\label{figPhaseAperture} (color online) Annular aperture placed over aberrated electron phase, showing an approximate $2\pi$ phase variation around the annulus. Aberrations are determined by eq. \ref{sawtoothfourierterms}, up to $5^{th}$ order. Outer aperture radius is $8.3$~mrad, inner aperture radius is $5.7$~mrad.}
 \end{figure}

The first five terms of this series are thus:
\begin{align}
\chi_{n}=&2\sin(\phi)-\sin(2 \phi)+\frac{2}{3}\sin(3 \phi) \nonumber\\
& \qquad \;  -\frac{1}{2}\sin(4 \phi)+\frac{2}{5}\sin(5 \phi) \label{sawtoothfourierterms}
\end{align}

By comparing equations \ref{aberchi} and \ref{sawtoothfourierterms}, we can see that if the $\theta$ dependence of $\chi$ can be removed, by applying an annular aperture,  and minimising all aberrations other than the $A_i$ values and their associated phase shifts, $\phi_{jj}$, the phase front can indeed be manipulated towards the ideal vortex phase structure.

 \begin{figure*}
\subfloat[Theoretical intensity and phase in the far--field. Intensity is represented by brightness and phase by hue, as indicated by the key.]{\label{figphplotSCALE}\includegraphics[width=0.38\linewidth]{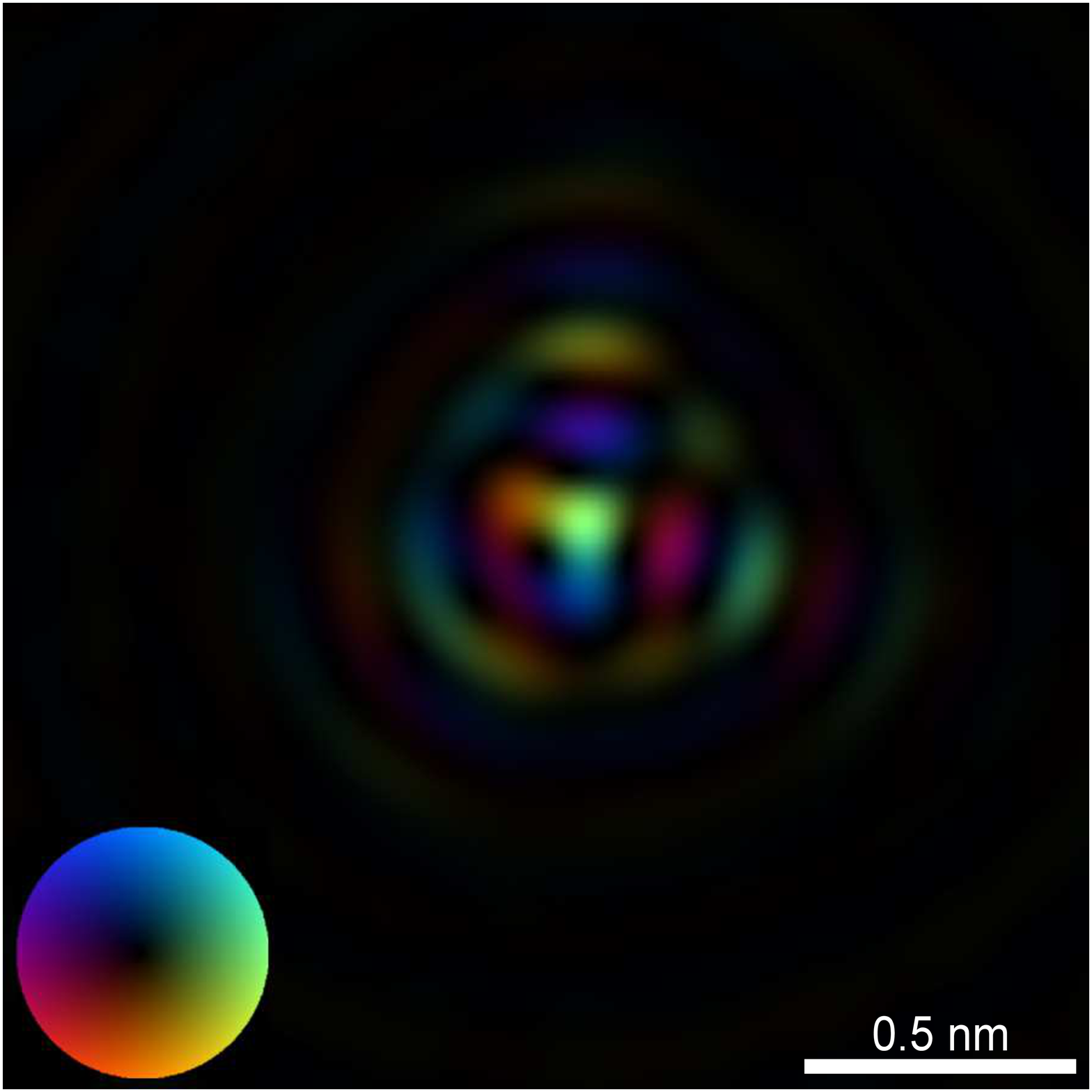}} 
\qquad \qquad
\subfloat[Experimental intensity pattern of aberrated probe and annular aperture system, in the sample plane.]{\label{expint}\includegraphics[width=0.38\linewidth]{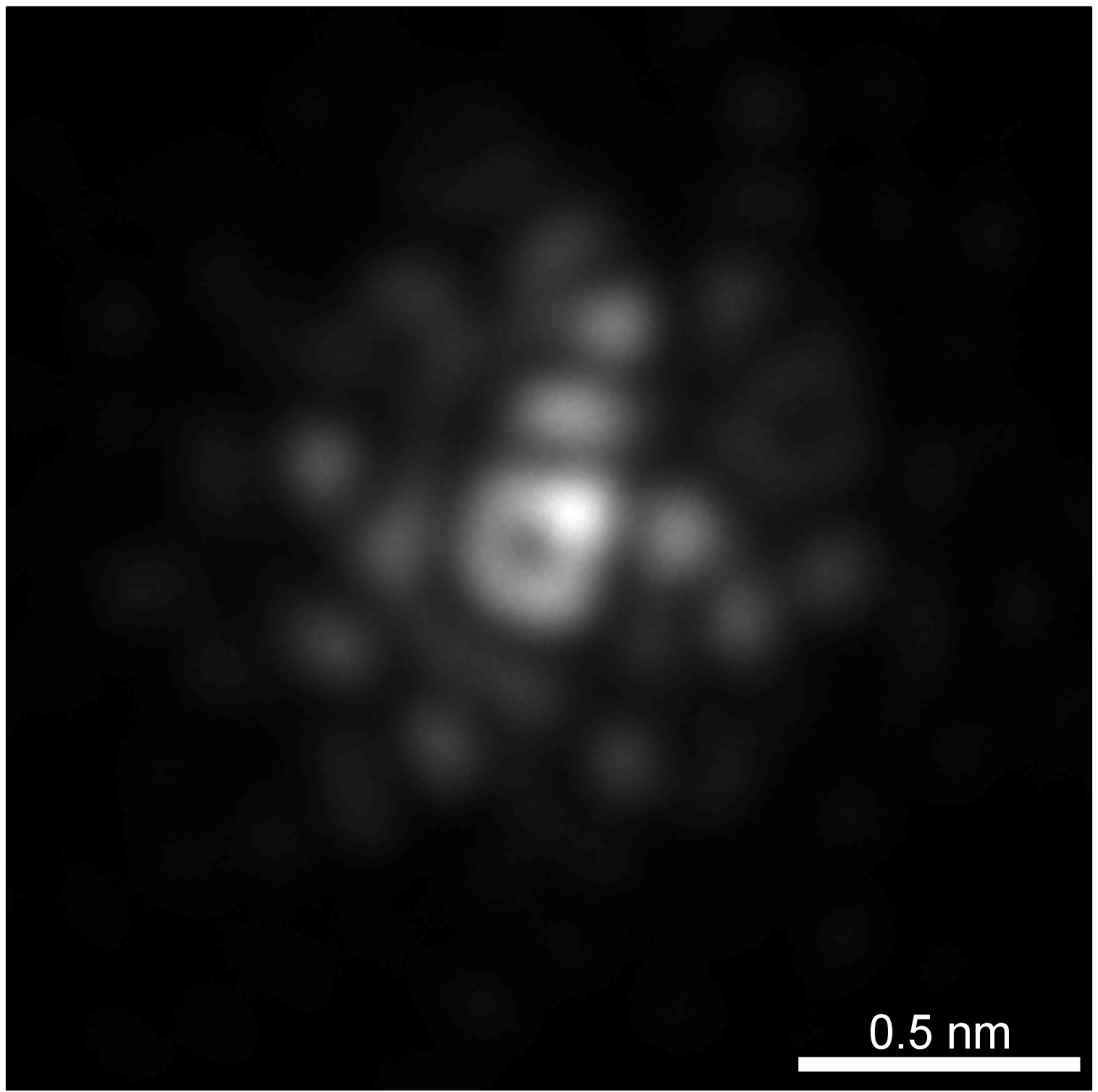}}
\caption{(color online) Theoretical and experimental comparison.}
 \end{figure*}

The required values of the $A_i$'s are each dependent on the value of $\theta$ selected by the annulus.
The angular size of the aperture should thus be selected to enable as many orders of $A_i$ to fall within achievable limits of the corrector.
Our simulations suggest that the positioning of the aperture can be displaced by up to $10\%$ of the radius laterally before the resulting beam is dramatically affected.
The effect of the number of Fourier terms included on the resulting vortex is demonstrated in detail in the Supplemental Material~\cite{suppmat}.
This set-up is illustrated in figure \ref{figPhaseAperture}, where Gibbs phenomenon features are visible in the upper quadrants of the annulus~\cite{jeffreys}, and the phases have been wrapped onto the $[-\pi, \pi]$ interval.

To perform this experiment, we used a double aberration corrected FEI Titan$^3$ TEM. The aberrations in the plane of the image forming lens were minimised.
The probe aberrations are adjusted using the probe corrector, and the aperture (with additional support bars added to the annulus) is inserted in the condenser plane. The aperture has an outer radius of $8.3$~mrad, which in our set-up lead to the desired values of $A_0=0.0896\,\mathrm{nm}$, $A_1=12.8\, \mathrm{nm}$ and $A_2=1.83\, \mathrm{\mu m}$ (which scale with $\theta$, $\theta^2$ and $\theta^3$ respectively) being within obtainable limits.
Their respective phase shifts are accordingly, $\phi_{11}=\nicefrac{3 \pi}{2}$, $\phi_{22}=\nicefrac{\pi}{4}$, and $\phi_{33}=\nicefrac{\pi}{2}$, for an $\ell=+1$ vortex.
 $A_3$--$A_5$ were minimised, as the desired values for these increase yet more strongly with $\theta$, and thus are beyond the achievable limits of the corrector.

This experimental set-up means that the far--field projection (equivalent to a 2D Fourier transform) of the annulus/phase system is incident on the sample plane, ideal for use in further experiments. 
The annular aperture forces the intensity pattern to become Bessel--like \cite{Durnin1, salehteich}, while propagation of the phase structure leads to a vortical wave field, with an $\ell=+1$ vortex remaining on axis.
This is demonstrated in figure \ref{figphplotSCALE}, where the characteristic rings of a Bessel function are visible, and furthermore, the phase is seen to vary from $0$ to $2\pi$ in an approximately linear fashion around the central zero.
The phase profile approximates the ideal vortex phase profile, at the region of highest intensity and thus such a set-up should result in a good approximation to an $\ell=+1$ electron Bessel beam being incident on the sample plane of a TEM.

The coupling of aberrations in the microscope, and parasitic aberrations present when the astigmatism parameters are increased result in other aberrations being non--zero during the experiment, accounting for differences between the idealised system (figure \ref{figphplotSCALE}), and the experimental result presented in figure \ref{expint}
\footnote{We note that the threefold astigmatism ($A_2$) employed experimentally had to be adjusted to half the theoretical value to enable a good match of resultant intensity profiles. This is likely due to an inconsistency in the definition of $A_2$ in the software with respect to equation \ref{aberchi}.}.

The level of the approximation to a certain vortex state can be quantified by a decomposition of the OAM modes in the superposition, to find the relative weighting of each mode.
We perform this analysis, following the work of Molina--Terriza \emph{et al.} and Berkhout \emph{et al.}~\cite{MolinaTerriza2002, BerkhoutMeasuringOAM} by projecting onto a spiral harmonic basis set, $\exp(\mathrm{i} \ell \phi)$.
In an ideal $\ell=1$, on--axis vortex state, the decomposition of modes would give 100\% weighting in the $\ell=1$ state.
However, it is firmly established that anisotropy, additional vortices and non--ideal vortex states significantly affect the decomposition, causing the distribution of modes to broaden~\cite{BerkhoutAnisotropy, MolinaTerriza2002, Gibson}.

\begin{figure}
 \includegraphics[width=0.99\linewidth]{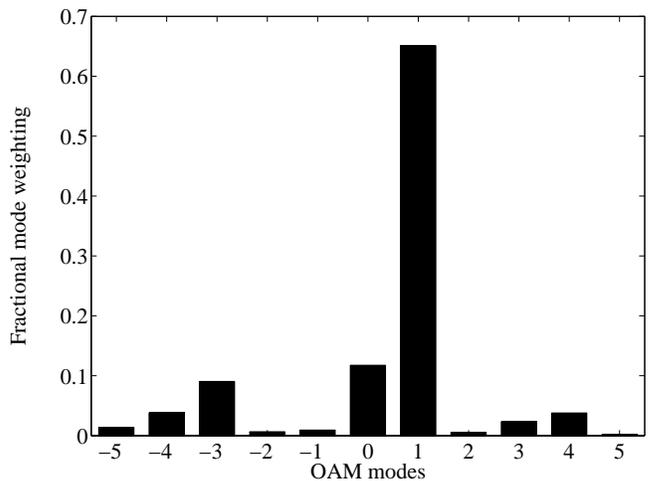}%
 \caption{\label{figOAMdecompCOMP} OAM decomposition of the far--field pattern of a $3^{rd}$ order corrected setup, within a ring of diameter $1$nm.}
 \end{figure}

We perform a mode decomposition, displayed in figure \ref{figOAMdecompCOMP}, of the Fourier series case, up to $A_2$, centred on the vortex core found in the centre of the doughnut intensity ring.
There is a clear peak in the $\ell=1$ mode.

Only the transmitted beam intensity falling within a disc of $1$nm diameter around the central vortex core is considered in this decomposition.
We find that approximately $50\%$ of transmitted beam intensity  is propagated into this region, of which $65\%$ is in the $\ell=1$ mode.
Overall this leads to $32\%$ of the transmitted beam being both around the vortex core, and in the desired mode.

For comparison, we perform the same analysis on the vortex produced by the holographic fork aperture.
The fork aperture is the binarised result of performing computer generated holography (CGH), with a desired (vortex) beam, and a reference (plane) wave \cite{bazhenovscrew, Goodman}. Illuminating the resulting aperture leads to a one--dimensional vortex array in the far--field \cite{Heckenberg1992, VerbeeckNature}.
In the fork aperture case, $17\%$ of transmitted intensity results in the region surrounding the $\ell=1$ vortex core, with $98\%$ of this being in the $\ell=1$ state.
Thus the fork aperture setup leads to approximately $17\%$ of the transmitted beam being both around the vortex core, and in the desired mode.

Both apertures transmit approximately the same fraction of incident electrons ($48\%$ and $50\%$ for the annular and fork apertures respectively), so it can be seen that the annulus technique presented here is transmitting almost twice the intensity of electrons in the desired OAM state into the region surrounding the vortex core than holographic fork aperture technique.
The mode decomposition of the annular setup can be increased to around $80\%$ in the $\ell=1$ state by considering a smaller disc, at the expense of electron intensity.

From these values, it can be seen that this new method of electron vortex beam production, leads to a significant increase in intensity in the $\ell=1$ mode surrounding the vortex compared to the holographic mask method, while additionally avoiding high intensity in other vortex modes, which is a key drawback of the holographic mask technique.

We must also note here that due to symmetries in the beam structure, we cannot determine from a single intensity pattern whether the produced beam is $\ell=+1$ or $\ell=-1$.
This is because the angle of twofold astigmatism ($\phi_{22}$) is varied freely during the experiment, and it is the relationship between the $\phi_{jj}$ which determine the vortex chirality. This uncertainty is typical of vortex work, as the beams of opposite chirality have no difference in intensity pattern.
Methods do exist to differentiate such beams~\cite{Giulio, maziluLP}, however, these require more complex experimental setups.

 \begin{figure}
 \includegraphics[width=0.99\linewidth]{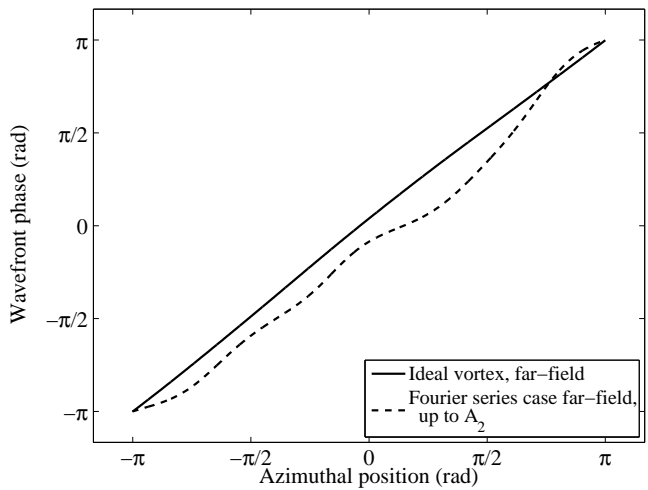}%
 \caption{\label{figPhaseProfile} Phase variation around the vortex core of an ideal vortex and the Fourier series system.}
 \end{figure}

The distribution of modes in figure \ref{figOAMdecompCOMP} can be understood by considering figure \ref{figPhaseProfile}.
This graph shows the variation in electron phase while completing a closed loop around the doughnut intensity profile, of both the far--field of the Fourier series method, and an ideal $\ell=1$ vortex.
The phase was measured around a circle centred on axis, at the radius of highest intensity.
The ideal case shows the vortex phase behaviour, while the Fourier series case shows small deviations from the ideal situation, with the variations in phase gradient contributing to the non $|\ell |=1$ mode decomposition. 

In this work we have designed and demonstrated a new, practical method for production of electron vortex beams in a TEM.
It has been shown that this system may be optimised such that up to $50\%$ of transmitted intensity is found around the vortex core, of which $65\%$ is in the desired $|\ell |=1$ mode.
The electron vortex produced in this way is therefore a less pure state than the vortex beams produced with the holographic mask method, but has a higher overall intensity of the desired mode in the vortex ring, and avoids  the conjugate and higher order beams produced in the holographic method. 

\begin{acknowledgments}
L.C., A.B., G.G. and J.V. acknowledge funding from the European Research Council under the 7th Framework Program (FP7), ERC Starting Grant No. 278510 - VORTEX.
A.B., A.L. and J.V. acknowledge financial support from the European Union under the 7th Framework Program (FP7) under a contract for an Integrated Infrastructure Initiative (Reference No. 312483 ESTEEM2).
J.V. was supported by funding from the European Research Council under the 7th Framework Program (FP7), ERC grant No. 246791 - COUNTATOMS.
R.V.B. acknowledges support from an FWO PhD fellowship grant (Aspirant Fonds Wetenschappelijk Onderzoek - Vlaanderen).
\end{acknowledgments}

\bibliography{Clark-AberrationVortexBib}
\end{document}